\PassOptionsToPackage{twoside=false}{geometry}

\documentclass[acmsmall,11pt,nonacm]{acmart}

\usepackage{graphicx}%
\usepackage{xcolor}%

\hypersetup{%
  colorlinks=true,
  linkcolor=blue,
  citecolor=blue,
  urlcolor=blue,
}

\begin{document}

\title[Counting and enumerating set partitions]{Methods for Counting and Enumerating Set Partitions}

\author{Arnav Khinvasara}
%\authornote{Long affiliations: University of California, Berkeley, USA; Unatech GmbH, Berlin, Germany.}
\affiliation{%
  \institution{University of California Berkeley}
  \department{Department of Electrical Engineering and Computer Sciences}
  \city{Berkeley}
  \state{CA}
  \country{USA}%
}
\affiliation{%
  \institution{Unatech GmbH}
%  \department{Research and Development}
%  \city{Berlin}
  \country{Germany}%
}

\author{Alexander Pikovski}
\authornotemark[1]
%\authornote{Long affiliation: Unatech GmbH, Berlin, Germany.}
\affiliation{%
  \institution{Unatech GmbH}
  \department{Research and Development}
  \city{Berlin}
  \country{Germany}%
}
\email{ap@unatech.de}

\begin{abstract}

Set partitions are arrangements of distinct objects into groups. 
After a brief review of the subject, we consider the task of counting and enumerating set partitions. The number of set partitions, known as Bell number, 
is a rapidly increasing number and does not have an explicit formula. We study approximate expressions for the Bell number given in the literature. We find that
an asymptotic formula of Moser and Wyman gives a surprisingly accurate approximation to the Bell number even for small set sizes. Furthermore, a simple expression due to Berend and Tasssa can be conveniently used to approximate the Bell number for small set sizes.
Next, we consider enumeration of set partitions.
The problem of listing all set partitions arises in a variety of settings, in particular in combinatorial optimization tasks. 
Algorithms for enumerating all set partitions are reviewed. The focus is on non-recursive algorithms without Gray code constructions. We compare the classic algorithm of Hutchinson with three more modern ones. Empirically, it is found that all of them scale exponentially with the set size. While the exact compiler and optimization settings do matter, it can be concluded that the algorithm of Djokic et al.\ is the fastest one, thus it is recommended for practical use.
\end{abstract}


\maketitle

\section{Introduction}

	A set partition is an arrangement of $n$ distinct objects into groups. 
	More formally, a partition of a set is a collection of non-overlapping sets (called groups or blocks)
	whose union is the original set. For example, the set $\{1,2,3\}$ has in total 5 partitions:
	%$\{\{1\},\{2\},\{3\}\}, 
	%\{\{1, 2\}, \{3\}\}, 
	%\{\{1, 3\}, \{2\}\}, 
	%\{\{1\} , \{2,3\}\}, 
	%\{\{1, 2, 3\}\}
	%$
	%
	$(\{1\},\{2\},\{3\})$, 
	$(\{1, 2\}, \{3\})$, 
	$(\{1, 3\}, \{2\})$, 
	$(\{1\} , \{2,3 \})$, 
	$(\{1, 2, 3\})$.

	In practical problems, set partitions arise in a variety of settings. 
	Usually one wants to find an optimal way to distribute $n$ objects into groups according to certain criteria.
	If $n$ is not too large, one can enumerate all possible set partitions and use a full search to find the optimal one.
	The total number of set partitions is a rapidly increasing number (see discussion in Sec.~\ref{sec:Bell}),
	and already for $n=17$ we have about $8 \cdot 10^{10}$ partitions.
	Thus, for $n$ larger than 17 or 18 (depending on the computational resources) listing all partitions is not feasible. 
	Still, it is a problem of practical importance to list all set partitions for a given $n$, if it is not too large.
	
	After briefly reviewing the application of set partitions, we discuss the number of set partitions (Bell numbers) and approximate formulas to calculate them.
	Then, we discuss and benchmark four algorithms for generating all set partitions, a classical one and three more modern ones. 
	We conclude by recommending the algorithm with the best performance and give an outlook to topics outside the scope of this study. 
	
	The problem of enumerating set partitions in hardware, as opposed to software, is considered in \cite{Butler-2015}.
	
	A related problem is to arrange $n$ distinct objects into groups size $k$. These are called restricted set partitions,
	in contrast to unrestricted set partitions which are the topic of the paper.

\section{How Set Partitions Arise in Practice}
\label{sec:Examples}

A common problem which leads to set partitions is: What are the placements of $n$ distinct items in at most $n$ boxes? 
	Similarly, set partitions arise in questions of the following type:
	What are the possible ways to serve $n$ dishes on plates?
	What are the possible ways to sit $n$ people at tables? 
	What are the possible ``rhyme schemes'' for $n$ lines? (E.~g.\ for 3 lines we have
	the 5 patterns 
	$a b c$,
	$a a b$,
	$a b a$,
	$a b b$,
	$a a a$.)
	
	Determining all set partitions has practical relevance in the setting of an optimization problem. Imagine you have $n$ different items, which need to be packed and shipped. Each parcel may contain one or more items. There are certain constraints (size, weight, ...) which are case-specific. The way to go about solving this problem algorithmically is to check all possible ways to pack the items, calculating the case-specific constraints for each one, and then choose the optimal one.
	
	Further applications of set partitions, among others to scheduling problems, are discussed in \cite{Hankin-2008}.

% Card shuffles ? 
%   https://www.whitman.edu/mathematics/cgt_online/book/section01.04.html

% Growth columns / partitions without two consecutive integers

\section{Number of set partitions (Bell numbers)}
\label{sec:Bell}

The number of set partitions of a set of size $n$ is called the Bell number $B_n$.
	The mathematical properties of Bell numbers are well-studied, see e.~g.\ \cite{Rota-1964}.
	To calculate Bell numbers, one can use for example the following recursion 
	(for further recursions, see \cite{Spivey-2008}):
	\begin{equation}
		B_{n} = 1 + \sum_{k=1}^{n-1} \binom{n-1}{k} B_k  .
		\label{eq:rec}
	\end{equation}

	This formula may be illustrated using the example of packing $n$ items in boxes as follows. Look at one specific item. This item goes either alone in a box, or with $k$ items. There are $\binom{n-1}{k}$ choices to find $k$ neighbours for this item, and the remaining $n - k - 1$ items can be packed in $B_{n-k-1}$ ways in other boxes, for $k = 0 \ldots n-2$; the case $k=n-1$ (all items on one box) in addition contributes one. 
	Thus in total, we have $\sum_{k=0}^{n-2} \binom{n-1}{k} B_{n-k-1} + 1$ ways to pack $n$ items, which, after simplification, is the same as Eq.~\ref{eq:rec}.
	
	The exact value for $B_n$ (see Table~\ref{table:1} for values up to $n=20$) can obtained with computer-algebra packages.
	We used SymPy and Matlab for numerical calculations.
	
	We now discuss several approximate expressions for the Bell numbers. 
	A large variety of different asymptotic and approximate expressions are described in the literature \citep{Knuth-book, Mansour-book, Grunwald-2025}. 
	We reviewed the different expressions and compared them numerically with exact values, to determine the most useful ones for practical use.
	Remarkably, we found that several asymptotic expressions for $B_n$ give a very good approximation even at small values of $n$. The best one is the following.
	
	The unique solution to the equation $\eta e^\eta = x$ (for positive $x$), is conventionally called Lambert's W function and denoted by $\eta = W(x)$.
	This function is not very common, but it is present in most numerical and symbolic libraries. 
	The expression for Bell numbers obtained by Moser and Wyman~\cite{Moser-1955} reads:
	
%	\begin{multline}
	\begin{equation}
		B_n \approx B_n^\ast = 
		\frac{e^{n W(n)+ n/W(n) - n - 1}}{\sqrt{W(n) + 1}} 
%		\\ \times 
		\left( 1 - \frac{W(n)^2 (2 W(n)^2+ 7 W(n) + 10)}{24 n (W(n)+1)^3} \right) .
		\label{MW}	
	\end{equation}
%	\end{multline}
	
	%
	%B_n \approx B_n^\ast = 
	%\frac{e^{n W + n/W - n - 1}}{\sqrt{W + 1}} 
	%\left( 1 - \frac{W^2 (2 W^2+ 7 W + 10)}{24 n (W+1)^3} \right)
%\end{multicols}

	While is an asymptotic expression for $B_n$, remarkably, it gives a very good approximation for all values of $n$.
	Specifically, comparing the result from Eq.~\ref{MW} with the exact values numerically, we found that the relative difference $|B^\ast_n - B_n|/B_n$ is smaller than $3\cdot 10^{-3}$ for all $2 \le n \le 50$. Moreover, for $n = 1 \ldots 7$, the integer part of $B_n^\ast$ is exactly equal to $B_n$.

\begin{table}[H]
\linespread{1.5}
\centering
\captionsetup{labelfont={bf},font={it,scriptsize}}
\caption{Comparing exact number of set partitions (Bell numbers) with the approximate expressions of Moser--Wyman (Eq.~\ref{MW}) and Berend--Tassa (Eq.~\ref{BT}).}
\scriptsize
%\rowcolors{1}{mygray}{mygray} 
%\begin{tabularx}{\columnwidth}{X X X X}
\begin{tabular}{l | l l l}
	\bottomrule
	\textbf{n} & \textbf{Set partitions $B_n$ (Exact number)} & \textbf{Moser--Wyman Approximation} & \textbf{Berend--Tassa Approximation} \\
	\hline
	1  & 1                    & 1                    & 1 \\
	2  & 2                    & 2                    & 2 \\
	3  & 5                    & 5                    & 5 \\
	4  & 15                   & 15                   & 15 \\
	5  & 52                   & 52                   & 52 \\
	6  & 203                  & 203                  & 212 \\
	7  & 877                  & 877                  & 957 \\
	8  & 4\,140               & 4\,143               & 4\,781 \\
	9  & 21\,147              & 21\,161              & 26\,107 \\
	10 & 115\,975             & 116\,040             & 154\,507 \\
	11 & 678\,570             & 678.9\,$\cdot 10^{3\phantom{1}}$ & 984\phantom{.}\,$\cdot 10^{3\phantom{1}}$ \\
	12 & 4\,213\,597          & 4.215\,$\cdot 10^{6\phantom{1}}$ & 6.70\,$\cdot 10^{6\phantom{1}}$ \\
	13 & 27\,644\,437         & 27.65\,$\cdot 10^{6\phantom{1}}$ & 48.5\,$\cdot 10^{6\phantom{1}}$ \\
	14 & 190\,899\,322        & 191.0\,$\cdot 10^{6\phantom{1}}$ & 372\phantom{.}\,$\cdot 10^{6\phantom{1}}$ \\
	15 & 1\,382\,958\,545     & 1.383\,$\cdot 10^{9\phantom{1}}$ & 3.01\,$\cdot 10^{9\phantom{1}}$ \\
	16 & 10\,480\,142\,147    & 10.48\,$\cdot 10^{9\phantom{1}}$ & 25.6\,$\cdot 10^{9\phantom{1}}$ \\
	17 & 82\,864\,869\,804    & 82.88\,$\cdot 10^{9\phantom{1}}$ & 229\phantom{.}\,$\cdot 10^{9\phantom{1}}$ \\
	18 & 682\,076\,806\,159   & 682.2\,$\cdot 10^{9\phantom{1}}$ & 2.14\,$\cdot 10^{12}$ \\
	19 & 5\,832\,742\,205\,057 & 5.834\,$\cdot 10^{12}$ & 20.8\,$\cdot 10^{12}$ \\
	20 & 51\,724\,158\,235\,372 & 51.73\,$\cdot 10^{12}$ & 211\phantom{.}\,$\cdot 10^{12}$ \\
	\bottomrule
%\end{tabularx}
\end{tabular}
\label{table:1}
\end{table}
%\begin{multicols}{2}
%\raggedcolumns%??????

	Berend and Tassa~\cite{Berend-2010} proved the following upper bound on Bell numbers:
	\begin{equation}
		B_n < \overline{B}_n = \left( \frac{0.792 \, n}{\log(n+1)} \right)^n ,
		\label{BT}
	\end{equation}
	valid for any $n$. Comparisons show that this bound is very tight for small $n$, and becomes less good for larger $n$. 
	%(For $n \le 8 $ it deviates by up to $15\%$, and for $n=20$ it overestimates the exact value by a factor of 4.)
	The formula for $\overline{B}_n$ is much simpler than Eq.~\ref{MW} and can be used to estimate Bell numbers for not too large $n$.

	The results are summarized in Table \ref{table:1}, where we show the exact value for $B_n$ compared to the approximations of Moser--Wyman and Berend--Tassa. The conclusion is that for smaller $n$, the simple expression of Eq.~\ref{BT} gives an excellent approximation while for larger $n$ one should use Eq.~\ref{MW}.

\section{Algorithms for listing set partitions}
\label{sec:Algorithms}

Finding the optimal arrangement of objects requires, as discussed above, in some cases a listing of all partitions of a set. The problem of finding an efficient algorithm for enumerating set partitions was repeatedly treated in the literature. However, a comparison of the different proposals, with an eye towards practical use, is lacking.
	
	The recursion of Eq.~\ref{eq:rec} and similar ones suggest that the problem of generation of all set partitions can be solved recursively. While this is true, recursive algorithms in practice are less efficient than non-recursive (iterative) ones. The main reason is that repeated function calling creates an overhead which is 
	also very much system- and compiler-specific. It is clear that, generally, iterative algorithms outperform recursive ones. Thus we only benchmark the non-recursive algorithms.
	
	A class of algorithms for generating set partitions are those based on Gray codes. Here, a Gray code is a one-to-one mapping from $\{1, \ldots, B_n\}$ to the set partitions for set size $n$, having certain properties. In other words, we get an indexing scheme (of a special type) for set partitions. Gray-code base approaches often deal with both restricted and unrestricted set partitions~\cite{Kawano-2005}. However, we do not consider Gray code constructions here.
	
	The oldest algorithm for calculating set partitions is due to Hutchinson~\cite{Hutchinson-1963}.
	This classical algorithm is also presented in the recent volume of the influential book by Knuth~\cite{Knuth-book}. However, Hutchinson's algorithm is definitely not the most efficient one, and we do not recommend its use in practical calculations (see discussion in Sec.~\ref{sec:Benchmarking}).
	
	A fast algorithm for generating all set partitions was presented by Semba~\citep{Semba-1984}.
	Another fast algorithm was devised by Er~\citep{Er-1988} (incidentally, using ideas from Gray codes but without explicitly constructing them). Later Djokic et.~al.\ presented another fast algorithm for generating set partitions~\citep{Djokic-1989}.
	
	A pedagogical exposition of the mentioned algorithms is given in \cite{Stamatelatos-2021}.

\section{Benchmarking of algorithms}
\label{sec:Benchmarking}

Since there is no clear theoretical analysis of the running time of algorithms for set partitions, we performed an empirical benchmarking. 
	(The empirical studies presented in \cite{Er-1988, Djokic-1989} are clearly out of date.)
	
	We have tested the algorithms
	%\cite{Hutchinson-1963}, \cite{Semba-1984}, \cite{Er-1988} and \cite{Djokic-1989}.
	\cite{Hutchinson-1963, Semba-1984, Er-1988, Djokic-1989}.
	The algorithms were tested both on Linux and on Windows operating systems. The code was written in C, to ensure lightweight binaries and fast running times. The compilers used were GNU C Compiler (GCC), Intel C Compiler (ICC) and Microsoft Visual C (MSVC). 
	
	Comparing different compilers on the same machine, we get the following pattern.
	For similar optimization levels, Hutchinson's algorithm runs significantly faster on GCC compiler than using the ICC compiler. For Semba's and Er's algorithm, GCC is faster than ICC, while for Djokic's the GCC is much slower than ICC.
	
	Comparing different operating systems (and compilers) on the same machine, we find that Linux-based code (GCC, ICC) is faster than Windows-based (MSVC); the former can be up to a factor of 2 faster.
	
	The compiler optimization level plays a decisive role, however, the difference between optimization level 2 and 3 is not large. Turning on some additional hardware-specific optimizations does not always lead to faster code. 
	
\begin{figure}[H]
	\centering
	\includegraphics[width=0.8 \linewidth]{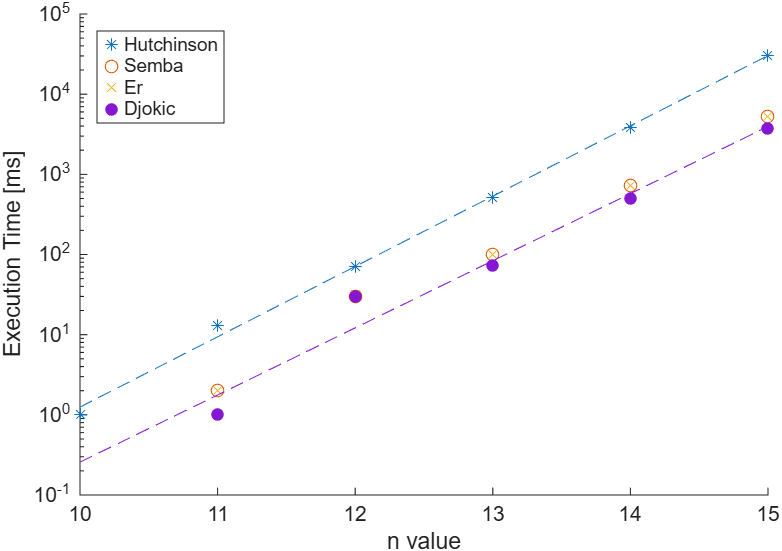}
	\caption{Execution time for the studied algorithms, for different $n$. Time is shown in logarithmic scale. Dashed lines are linear fits to the data points. ICC compiler with highest optimization level was used on Intel Core i7-11700 @ 2.50 GHz CPU.}
	\label{figure:1}
\end{figure}

	Now, comparing different algorithms with the same compiler (ICC), with highest compiler optimization. Results for running times are shown in Figure~\ref{figure:1}. Hutchinson's algorithm is significantly slower than the other three. The algorithms of Semba and Er are faster, and that of Djokic et al.\ is the fastest one. It is clearly seen that the running time for all four algorithms scales exponentially with $n$.

% === TIMING RESULTS (milliseconds) ===
% 
% n values:        1       2       3       4       5       6       7       8       9      10      11      12      13      14      15
% Hutchinson       0       0       0       0       0       0       0       0       1       1      13      70     510    3850   30512
% Semba            0       0       0       0       0       0       0       0       0       0       2      30     100     720    5285
% Er              0       0       0       0       0       0       0       0       0       0       2      32     100     720    5318
% Djokic           0       0       0       0       0       0       0       0       0       0       1      30      72     506    3746

%\begin{table}
%\centering
%\begin{tabular}{l|rrrrrrrrrrr}
%\textbf{$n$ value:}  &     \textbf{8}  &     \textbf{9}  &    \textbf{10}   &   \textbf{11}   &   \textbf{12}  &    \textbf{13}  &    \textbf{14}  &    \textbf{15} \\ \hline
%Hutchinson \cite{Hutchinson-1963}    		&     0  &     1  &     1   &   13   &   70  &   510  &  3850  & 30512 \\
%Semba \cite{Semba-1984}         		&     0  &     0  &     0   &    2   &   30  &   100  &   720  &  5285 \\
%Er \cite{Er-1988}            		&     0  &     0  &     0   &    2   &   32  &   100  &   720  &  5318 \\
%Djokic et al.\ \cite{Djokic-1989}        		&     0  &     0  &     0   &    1   &   30  &    72  &   506  &  3746
%\end{tabular}
%\caption{CPU run time (in milliseconds) for the studied algorithms, for different $n$. Presented tests are performed on a Linux desktop PC, with GCC compiler.}
%\label{t:cmp}
%\end{table}

For practical purposes, we recommend the algorithm of \cite{Djokic-1989}. 
	The algorithm is relatively short and easy to implement.
	Compiler optimization should be set at least at level 2. On Linux, the Intel Compiler (ICC) for this algorithm gives faster code than GNU C Compiler (GCC).

\section{Conclusions and outlook}
\label{sec:Conclusions}

After an introduction to set partitions and some general background, we have presented  useful approximation formulas for the calculation of the number of set partitions (Bell numbers), both for small and large set sizes. Three state-of-the-art nonrecursive algorithms~\citep{Semba-1984,Er-1988,Djokic-1989} for generating set partitions were benchmarked, with a comparison with Hutchinson's classical algorithm~\citep{Hutchinson-1963}. As a result, the algorithm of Djokic et al.~\cite{Djokic-1989} is recommended for practical use.

	A problem closely related to finding all set partitions is that of finding all partitions with a maximal block size $p$, known as restricted set partitions. Some of the algorithms for unrestricted set partitions can be modified to yield restricted set partitions. Together with a review and benchmarking of Gray-code based algorithms, this warrants a separate investigation. Furthermore, it would be useful to study parallelizable algorithms for listing set partitions.

% If you must keep the ACM visual styling but want chronological ordering, download ACM-Reference-Format.bst into your project folder, open it in a text editor, find and comment out (using %) any lines containing the SORT directive, and save it.
%\bibliographystyle{ACM-Reference-Format}
\bibliographystyle{unsrtnat}

\bibliography{set_partitions_bib}

\end{document}